# Application of Small-Angular Magnetooptic Polarimetry for Study of Magnetogyration in $(Ga_{0.3}In_{0.7})_2Se_3$ and $SiO_2$ Crystals

O. Krupych, Yu. Vasylkiv, D. Adamenko, R. Vlokh and O. Vlokh

Institute of Physical Optics, 23 Dragomanov St., 79005 Lviv, Ukraine,
e-mail: vlokh@ifo.lviv.ua

## Abstract

We present the results of studies for magnetogyration (MG) effect in $(Ga_{0.3}In_{0.7})_2Se_3$ and $SiO_2$ crystals performed with the small-angular polarimetric mapping technique. It is shown that the MG effect magnitude is comparable with the experimental error.

## 1. Introduction

Optical activity effects are divided into two types, following from their symmetry peculiarities and experimental manifestations. The effects of spatial dispersion, or gyration, belong to the first type described by the symmetry $\infty 2$, while the Faraday-type effects form the second one (the symmetry $\infty / m$). These effects can be experimentally separated using a reversal of wave vector direction. The latter would lead to doubling the optical Faraday rotation (FR) and compensating the optical gyration rotation.

The peculiarities under discussion may be proven on the basis of simple phenomenological analysis of the constitutive relations. The electric displacement of a plane electromagnetic wave propagating in a magnetically non-ordered medium ($D_i^w$), dependent on the optical frequency $w$, and the dielectric permittivity ($e_{ij}$) that accounts for the first-order spatial dispersion may be written respectively as

$$D_i^w = e_{ij}^0 E_j^w + g_{ijk} \frac{\partial E_j^w}{\partial x_k} = \\ = e_{ij}^0 E_j^w + i e_{ijl} g_{lk} k_k E_j^w , \qquad (1)$$

$$e_{ij} = e_{ij}^0 + i g_{ijk} k_k = e_{ij}^0 + i e_{ijl} g_{lk} k_k . \qquad (2)$$

At the same time, when external magnetic field is applied and the spatial dispersion is absent, one has

$$D_i^w = e_{ij}^0 E_j^w + i J_{ijk} H_k E_j^w = \\ = e_{ij}^0 E_j^w + i e_{ijl} a_{lk} H_k E_j^w , \qquad (3)$$

$$e_{ij} = e_{ij}^0 + i e_{ijl} a_{lk} H_k . \qquad (4)$$

As regards the notations used in Eqs. (1)–(4), $e_{ij}^0$ is a real part of the dielectric permittivity, $E_j^w$ the electric field of optical wave, $H_k$ the external magnetic field, $J_{ijk}$ and $g_{ijk}$ respectively the third-rank axial and polar tensors antisymmetric in $i$ and $j$ indices, $k_k$ the wave vector, $x_k$ the coordinate, $a_{lk}$ and $g_{lk}$ respectively the second-rank polar Faraday tensor and axial gyration tensor and $e_{ijl}$ the unit antisymmetric third-rank Levi-Civita pseudotensor. It is seen from the relations (1)–(4) that both the spatial dispersion and the magnetic field could lead to appearance of imaginary part in the dielectric permittivity. On the other hand, it follows from the Hermitian conditions that the imaginary parts of the dielectric permittivity $r_n \sim ie_{ijl} g_{lk} k_k$ and $r_n \sim ie_{ijl} a_{lk} H_k$ should be purely antisymmetric for transparent optical media. Antisymmetric part of a second-rank polar tensor is dual to an axial vector $r_n$. This means that the vector of electric displacement,

$$D_i^w = [r_n \times E_j^w], \qquad (5)$$

should rotate in such a medium, thus reflecting a vector product available in Eq. (5). A sign of the rotation of polarization plane ($r_n \sim g_{lk} k_k$) should depend either on the wave vector sign (in the case of spatial dispersion effect) or on the sign of magnetic field ($r_n \sim a_{lk} H_k$) for the Faraday optical activity.

However, the optical rotation effect observed in some pyroelectric crystals in the magnetic field, which has been explained initially as a 'magnetogyration' (MG) [1] and then as a combined influence of the electric and magnetic fields [2,3], has also been separated from the FR by reversing the wave vector. In relation to this problem, it is worth noticing that the MG as a spatial dispersion effect induced by magnetic field should not manifest itself as an optical rotation, since, according to the Onsager principle, it ought to lead to changes in the real part of the dielectric permittivity (or the refractive indices). This effect is known as a '$kH$-effect' [4]. In most of the previous studies [5–7], the MG has been separated from the FR by rotating samples by 180$^o$ around the axis perpendicular to the optic axis and, in such a way, changing the sign of the polar third-rank MG tensor $d'_{ikm}$ in the coordinate system related to $k_k$ and $H_m$ vectors. Under the operation described above, the MG rotation $r \sim d'_{ikm} k_k H_m$ should change its sign, contrary to the FR. In the works [5–7], the MG effect has been observed in such absorbing crystals as CdS, $(Ga_xIn_{1-x})_2Se_3$ (x=0.3 and 0.4) and $Bi_{12}GeO_{20}$.

However, our recent study [8] of the magnetooptic rotation (MOR) in CdS crystals, performed on the basis of small-angular polarimetric mapping, has testified the important experimental fact: there is no difference, within the experimental accuracy, between the FR values measured under the wave vector reversal, at least for the parameters averaged over the angle of laser beam divergence. Being able to separate the polarimetric data corresponding to the chosen directions of light propagation in crystal, we have demonstrated a clear advantage of small-angular magnetooptic polarimetric mapping, when compare with the single-ray polarimetry, in particular when the conditions of complicated experiments require sample position changes. We have shown in the mentioned work [8] that, within the limits of experimental accuracy, the MG effect does not exist in CdS crystals. The present paper is devoted to reinvestigation of MG effect in $(Ga_{0.3}In_{0.7})_2Se_3$ crystals belonging to the point group of symmetry 6, while using the same small-angular polarimetric mapping. We will also compare the results with those obtained also for $SiO_2$ crystals, which certainly should not possess the MG, due to general symmetry limitations.

## Experimental

A number of inevitable error sources exist in case of magnetooptic measurements. These are: (1) angular distribution of the FR near the optic axis, even within a small angle of a few angular degrees; (2) small divergence of the laser beam ($4\times10^{-3}\,rad$ or 0.23deg); (3) inhomogeneity of the magnetic field; (4) imperfections of crystalline samples. These sources of errors acquire a primary importance when one compares MORs obtained by probing the sample in the opposite directions, because the rotation of the latter by 180$^o$ around the axis perpendicular to the beam might lead to additional errors, due to misalignment of the sample orientation. Using a small-angular imaging polarimetric technique, it is possible to account for most of these errors. Of course, the accuracy of polarization measurements in frame of single-ray polarimetry is usually higher than that typical for the imaging polarimetry that operates with a wide light beam (notice that here we use the term ‘beam’ in the meaning of ‘bundle of rays’). Nonetheless, the accuracy of polarization azimuth orientation achieved in the present work is not worse than $4\times10^{-2}$ deg. Moreover, of a greater importance are misalignments in the propagation direction of the laser beam appearing under the rotation of sample by $180^{o}$. Really, they could yield different values of the FR, which might be then erroneously interpreted as a MG effect. The main problem that appears in the course of MG measurements is to obtain and analyze polarimetric data that refer to the same optical path in crystal (e.g., the optical path along the optic axis – see Fig. 1).

We used in our studies the imaging polarimetric setup presented in Fig. 1. The difference from a usual imaging polarimetry (see, e.g., [9]) consists in the use of conical probing beam, instead of a parallel one. In our experiments, the divergence angle of the conical beam was $3.49\times10^{-2}$ rad or approximately 2 deg. We used He-Ne laser ($l = 632.8nm$) as a source of optical radiation. Sample 8 was positioned at the beam waist. Objective lens 10 imaged the cross-section of the light beam passed analyzer 9 onto the sensor of CCD camera 11. The image obtained by the camera corresponded to the angular aperture of about $3.49\times10^{-2} rad$ (or $\sim 2\deg$). That is why this technique is called as ‘small-angle polarimetric mapping’. In order to avoid speckle structure of the obtained images, a coherence scrambler 4 was used. The angular divergence of the probing beam was limited by the dimensions of light channel in magnetic core 7.

The plane-parallel crystal sample was placed between the poles of electromagnet. The distance between the poles (54 mm) was large when compare to the sample thickness, thus allowing us to reduce inhomogeneities of the magnetic field (and the appearance of transverse component of that field) through the sample thickness to a negligibly small value. The estimated Cotton-Mouton birefringence induced by the transverse magnetic field was less than $\sim 10^{-8}$. The sample was positioned in the same manner as it was done in our recent studies [5,6], i.e. after aligning the centre of the conoscopic rings with the light beam centre.

The small-angular maps of the polarization azimuth were obtained without any magnetic field and for the case of direct magnetic field of $H_z = 4.63 kOe$, for the two orientations of sample (a 'direct' one and that corresponding to the sample rotated by 180°). Basing on these maps, we have calculated the MORs for the both wave vector directions (+*k* and –*k*).

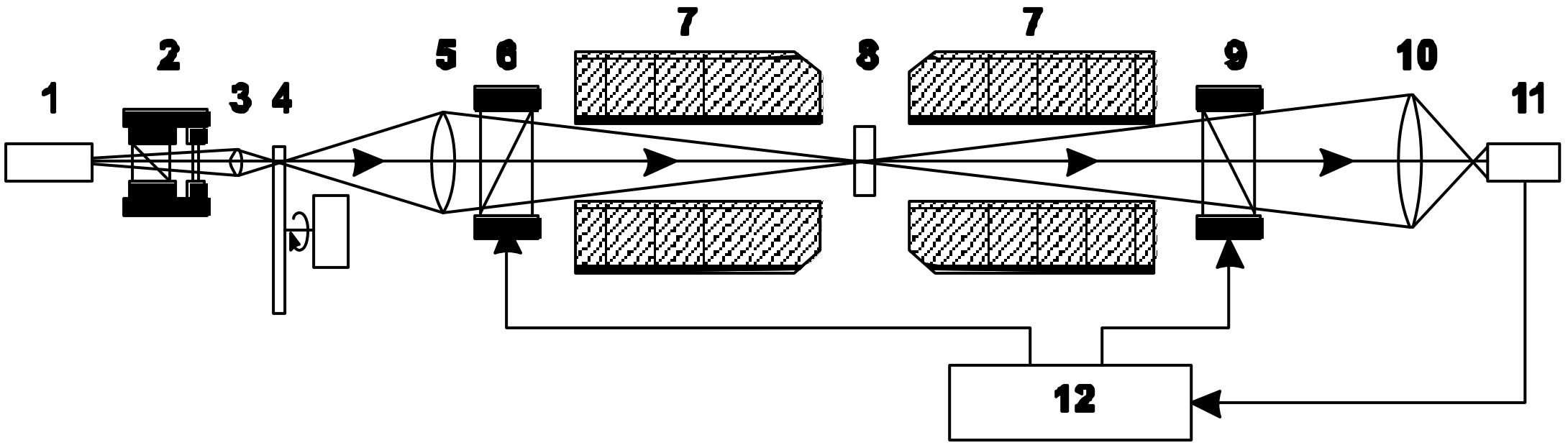

**Fig. 1.** Optical scheme for small-angle magnetooptic polarimetric mapping: 1 – laser; 2 – circular polarizer (linear polarizer and a quarter-wave plate); 3 – short-focus lens; 4 – coherence scrambler; 5 – long-focus lens; 6 – linear polarizer (Glan prism) with motorized rotary stage; 7 – magnetic core; 8 – sample; 9 – analyzer (Glan prism) with motorized rotary stage; 10 – objective lens; 11 – CCD camera; 12 – computer.

## Results and discussion

### 1. Quartz crystal

The small-angular maps of polarization azimuth for the quartz crystal are shown in Fig. 2. The MOR is a difference between the azimuths detected in the presence ($a_H$) and absence ($a_0$) of the magnetic field and so it could be calculated as

$$\Delta a = a_H - a_0 . \qquad (6)$$

The maps of that difference are shown in Fig. 3. It is necessary to analyze carefully these maps for correct determination of the MOR. A first important problem is identifying a location of optic axis outlet on the particular map. To solve this problem, we have simulated numerically the maps of both the polarization azimuth and the ellipticity across the conical light beam emergent from a crystalline quartz plate. The basic formulae for the ellipticity and the azimuth of polarization have been taken from [10]. We assume that the sample is cut perpendicular to the optic axis (the latter is parallel to crystallographic Z axis) and the incoming convergent beam is linearly polarized, with the cone axis being parallel to the optic axis. The maps calculated this way are presented in Fig. 4. It is evident that the polarization azimuth cannot be used as a criterion while identifying the optic axis, since its value varies only slightly across the map. The maximum deviation of the azimuth from the value corresponding to the optic axis does not exceed 0.12°, i.e. it is approximately equal to our experimental errors. On the other hand, one can clearly see from the map that the minimum value of the polarization ellipticity lies just in the centre of the pattern corresponding to the optic axis position, while the ellipticity deviation range reaches up to 2.5°. Therefore the position of the minimum on the polarization ellipticity map may be considered as a criterion, while identifying the optic axis.

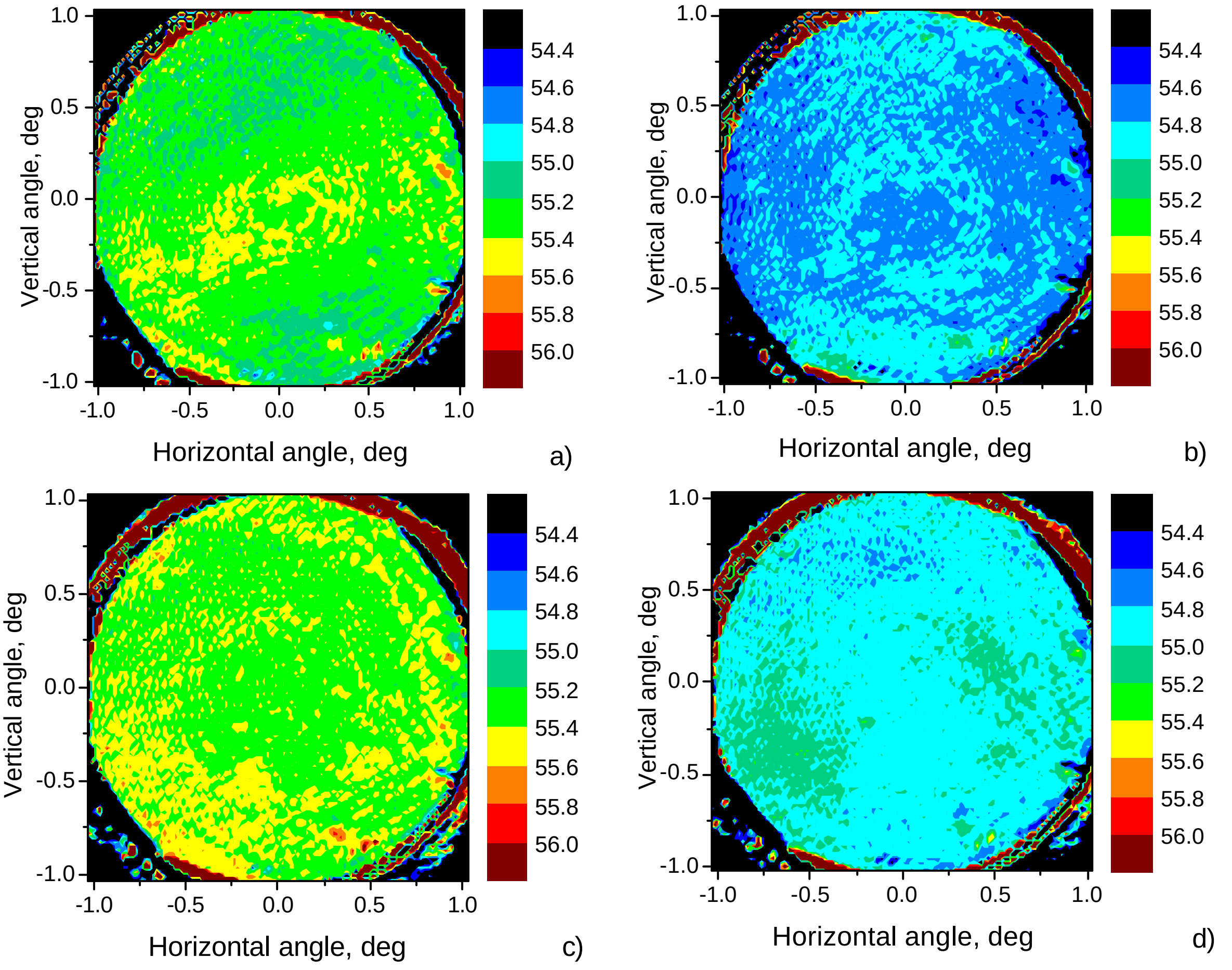

**Fig. 2.** Maps of polarization azimuth (in angular degrees) for the quartz crystal: (a) $+k$, $H = 0$, (b) $+k$, $H = 4.3$ Oe, (c) $-k$, $H = 0$ and (d) $-k$, $H = 4.3$ Oe.

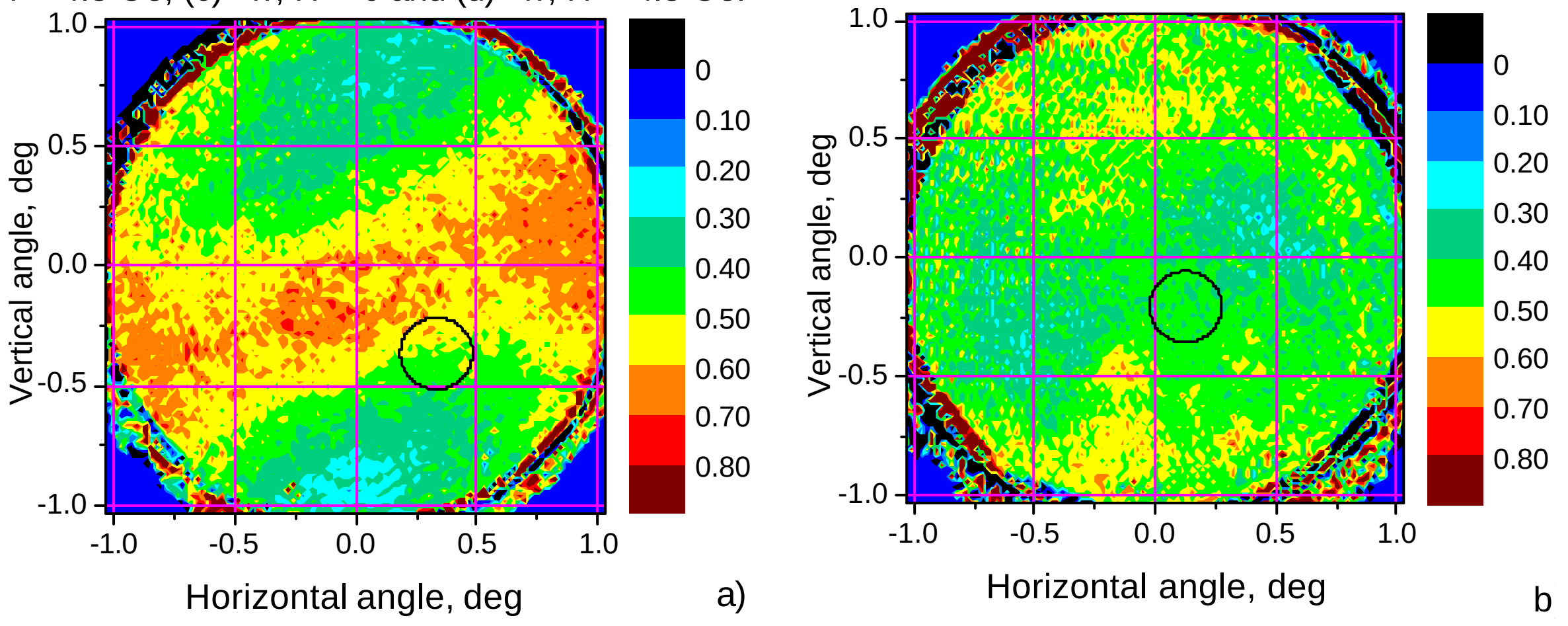

**Fig. 3.** Maps of MOR (in angular degrees) for the quartz crystal: (a) $+k$ and (b) $-k$.

Then we have analyzed the experimental maps of polarization ellipticity for the both directions of light propagation (see Fig. 5) obtained at a zero magnetic field and have identified the optic axis outlets. We have marked circular regions with the angular dimension of 0.23° = 4 mrad that correspond to a divergent laser beam used in single-ray polarimetry. The positions of these regions indicate to outlets of the optic axis. Next, we have calculated the mean values of the MOR for the chosen regions and the corresponding experimental accuracy (see the maps presented in Fig. 3). The following results are obtained for those regions (designated as circles in Fig. 3): $\Delta a^{+} = 0.495^{\circ} \pm 0.055^{\circ}$ and $\Delta a^{-} = 0.413^{\circ} \pm 0.044^{\circ}$.

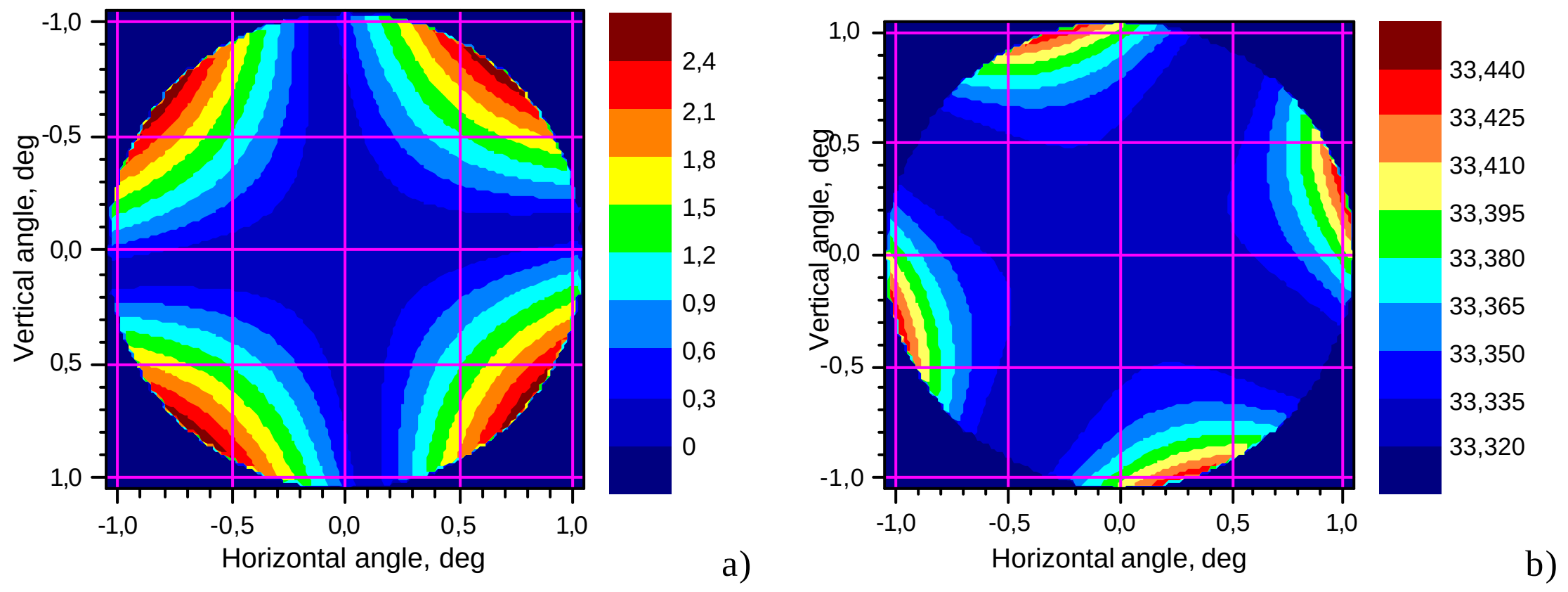

**Fig. 4.** Calculated maps of polarization ellipticity (a) and azimuth (b) for the light passed the quartz crystal. The azimuth of linearly polarized incoming light and the sample thickness are taken to be 45° and 4.25 mm, respectively.

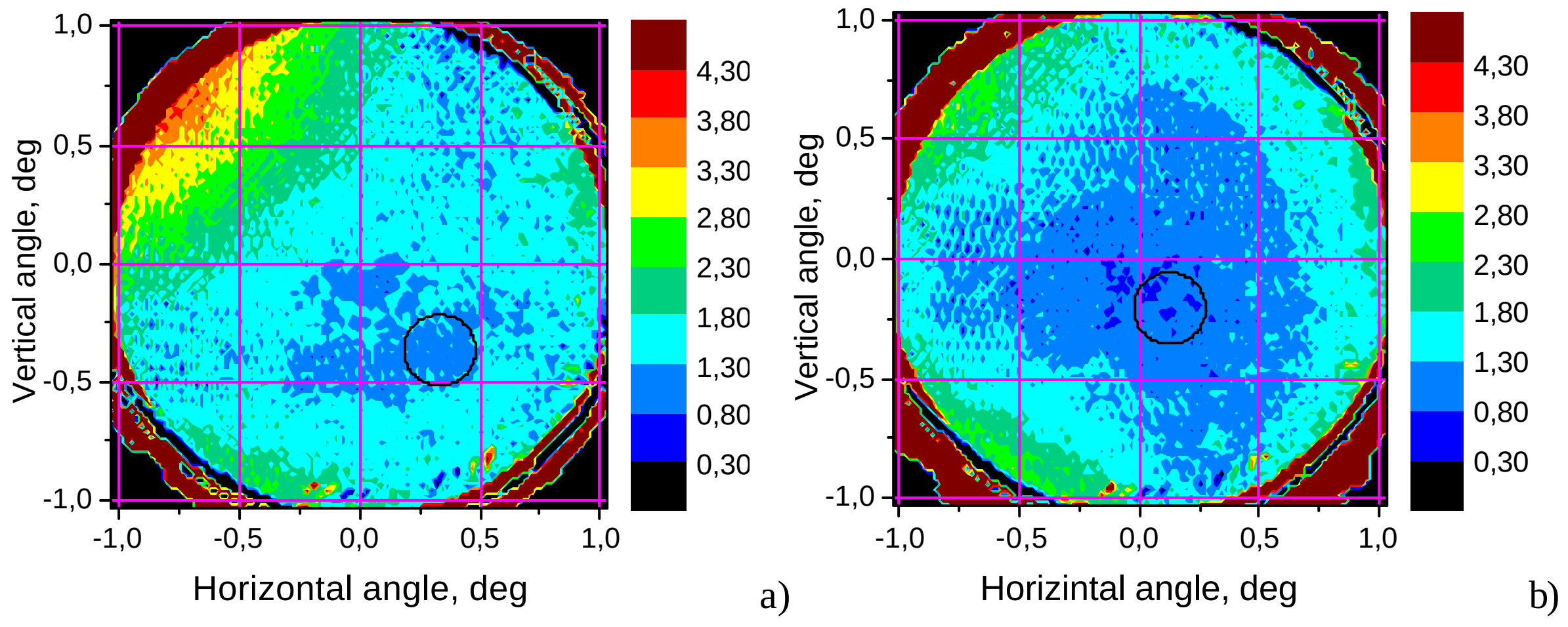

**Fig. 5.** Maps of light ellipticity (in angular degrees) for the quartz crystal: (a) +*k* and (b) –*k*.

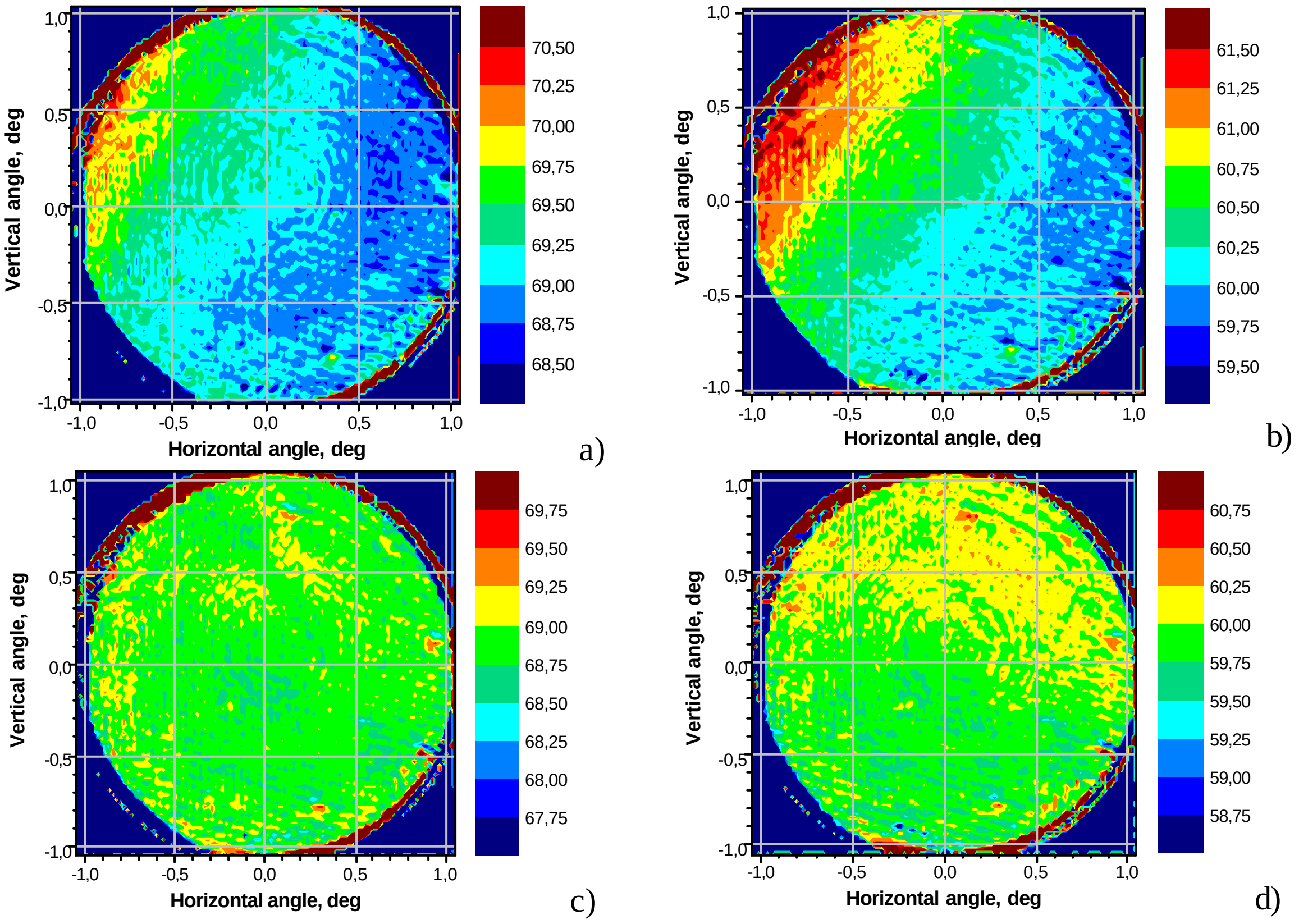

**Fig. 6.** Maps of polarization azimuth (in angular degrees) for $(Ga_{0.3}In_{0.7})_2Se_3$ crystal: (a) $+k$, $H$ = 0, (b) $+k$, $H$ = 4.3 Oe, (c) $–k$, $H$ = 0 and (d) $–k$, $H$ = 4.3 Oe.

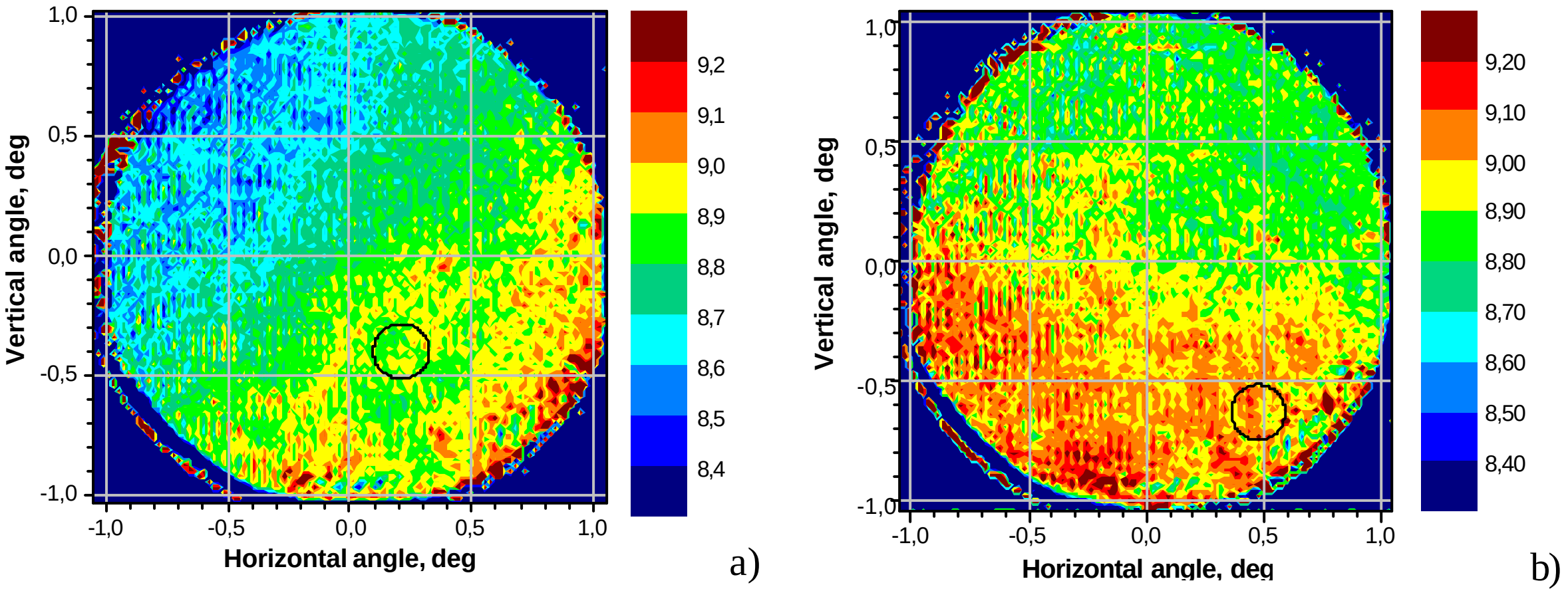

**Fig. 7.** Maps of MOR (in angular degrees) for $(Ga_{0.3}In_{0.7})_2Se_3$ crystal: (a) $+k$ and (b) $–k$.

Non-reciprocal MOR (abbreviated as NRMOR) is a difference of MORs ? α for the opposite directions of the wave vector (+*k* and –*k*):

$$d(\Delta a) = \Delta a^{+} - \Delta a^{-} . \qquad (7)$$

For the quartz crystal, the NRMOR calculated for the selected regions is $d(\Delta a) = 0.081^{\circ} \pm 0.049^{\circ}$. It is known that the NRMOR should be interpreted as a consequence of MG effect. However, the MG in quartz crystals is evidently forbidden by symmetry. This means that the NRMOR in quartz should be zero. As a result, the obtained value of the NRMOR might be reasonably considered as the total experimental error inherent in the measuring technique and the setup used in this work.

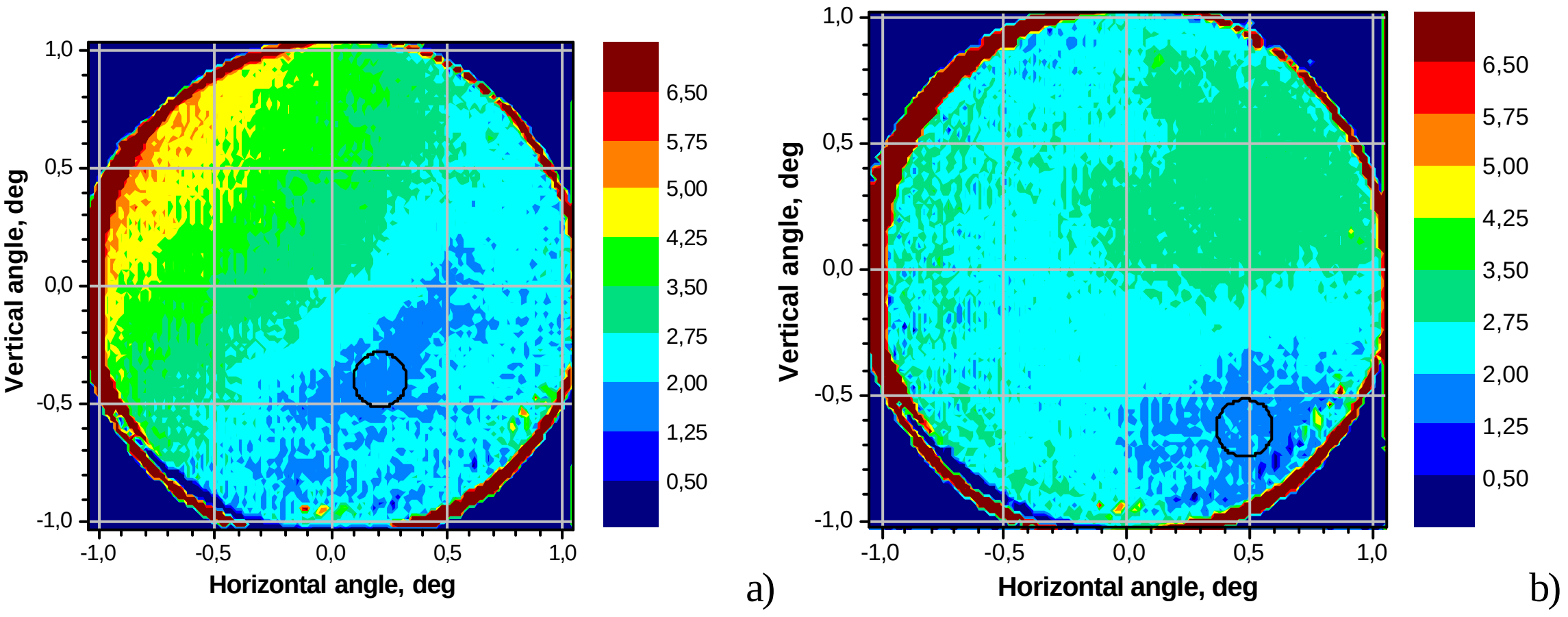

**Fig. 8.** Maps of light ellipticity (in angular degrees) for $(Ga_{0.3}In_{0.7})_2Se_3$ crystal: (a) $+k$ and (b) $-k$.

## 2. $(Ga_{0.3}In_{0.7})_2Se_3$ crystals

The small-angular maps of the polarization azimuth for $(Ga_{0.3}In_{0.7})_2Se_3$ crystal are shown in Fig. 6. The maps of the MOR are depicted in Fig. 7. After that, we have determined the location of the optic axis by means of technique similar to that described above and using the maps of polarization ellipticity (see Fig. 8). Then we have calculated the following mean values of the MOR for the chosen regions denoted by circles in Fig. 7: $\Delta a^+ = 8.897^\circ \pm 0.066^\circ$ and $\Delta a^- = 9.021^\circ \pm 0.090^\circ$. The NRMOR calculated for the selected regions is $d(\Delta a) = -0.124^\circ \pm 0.079^\circ$. The MG coefficient calculated from this value is equal to $d = (7.4 \pm 4.7) \cdot 10^{-11} Oe^{-1}$. Notice that the MG coefficient obtained by using single-ray polarimetric technique has been much larger ($d = 24.5 \times 10^{-11} Oe^{-1}$ [5,6]). The analysis performed in this work testifies that the mentioned results obtained previously should be incorrect. Moreover, our present results show that the NRMOR magnitude (the same refers to the MG one) is close to the experimental error (the relative error for the MG in $(Ga_{0.3}In_{0.7})_2Se_3$ crystal is 64%). Let us take into account that the relative error for the quartz crystals is approximately the same (60%) and the MG effect is forbidden in crystals of the point symmetry group 32 and for the experimental geometry used by us. Then one can conclude that the MG effect in the crystals under study does not exceed the level of experimental errors or does not exist at all.

## Conclusion

The results obtained in the present work demonstrate that the previously obtained data concerned with the observation of MG effect in $(Ga_{0.3}In_{0.7})_2Se_3$ crystals involve the errors, which might appear due to misalignments of sample under its rotation by 180° in the optical system and a small divergence of the laser beam.

## References

1. Zheludev I.S., Vlokh O.G. and Sergatyook V.A. Ferroelectrics **63** (1985) 97.
2. Vlokh O.G. and Sergatyook V.A. Izv. Akad. Nauk USSR **291** (1986) 832.
3. Vlokh O.G. and Sergatyook V.A Ferroelectrics **80** (1988) 313.
4. Krichevtsov B.B., Pisarev R.V., Rzhevsky A.A., Gridnev V.N. and Weber H.-J. Phys. Rev. B **57** (1998) 14611.

5. Vlokh R., Vlokh O. G., Klymiv I. and Adamenko D. Ukr. J. Phys. Opt. **3** (2002) 166.
6. Vlokh R., Vlokh O. G., Klymiv I. and Adamenko D. Ukr. J. Phys. Opt. **3** (2002) 271.
7. Vlokh R., Adamenko D., Say A., Klymiv I. and Vlokh O.G. Ukr. J. Phys. Opt. **5** (2004) 57.
8. Vlokh R.O., Adamenko D.I., Krupych O.M. and Vlokh O.G. Ferroelectrics (2007) at press.
9. Vlokh R., Krupych O., Kostyrko M., Netolya V. and Trach I. Ukr. J. Phys. Opt. **2** (2001) 154.
10. Konstantinova A.F., Grechushnikov B.N., Bokut B.V. and Valyashko E.G. Optical properties of crystals. Minsk, Nauka i Technika (1995). (in Russian).